\documentstyle[12pt,aaspp4]{article}

\newcommand{\etal}{{\it et al}}
\newcommand{\eg}{{\it eg}}

\begin{document}

\title{
	Caltech Faint Field Galaxy Redshift Survey IX:\\
	Source Detection and Photometry
	in the Hubble Deep Field Region\altaffilmark{1}
}
\author{
	David W. Hogg\altaffilmark{2,3,4,5},
	Michael A. Pahre\altaffilmark{3,5,6},
	Kurt L. Adelberger\altaffilmark{3},
	Roger Blandford\altaffilmark{2},
	Judith G. Cohen\altaffilmark{3},
	T. N. Gautier\altaffilmark{7},
	Thomas Jarrett\altaffilmark{8},
	Gerry Neugebauer\altaffilmark{3},
	Charles C. Steidel\altaffilmark{3}
}
\altaffiltext{1}{
Based on observations made at the Palomar Observatory, which is owned
and operated by the California Institute of Technology; and with the
NASA/ESA Hubble Space Telescope, which is operated by AURA under NASA
contract NAS 5-26555.
}
\altaffiltext{2}{
Theoretical Astrophysics, California Institute of Technology,
mail code 130-33, Pasadena CA 91125, USA
}
\altaffiltext{3}{
Palomar Observatory, California Institute of Technology,
mail code 105-24, Pasadena CA 91125, USA
}
\altaffiltext{4}{
Institute for Advanced Study,
Olden Lane, Princeton NJ 08540, USA;
{\tt hogg@ias.edu}
}
\altaffiltext{5}{
Hubble Fellow
}
\altaffiltext{6}{
Harvard-Smithsonian Center for Astrophysics, 60 Garden Street,
Mail Stop 20, Cambridge, MA 02139; {\tt mpahre@cfa.harvard.edu}
}
\altaffiltext{7}{
Jet Propulsion Laboratory, Mail Stop 264-767,
4800 Oak Grove Drive, Pasadena CA 91109, USA
}
\altaffiltext{8}{
Infrared Processing and Analysis Center, California Institute of Technology,
mail code IPAC 100-22, Pasadena CA 91125, USA
}

\begin{abstract}
Detection and photometry of sources in the $U_n$, $G$, $\cal R$, and
$K_s$ bands in a $9\times 9$~arcmin$^2$ region of the sky, centered on
the Hubble Deep Field, are described.  The data permit construction of
complete photometric catalogs to roughly $U_n=25$, $G=26$, ${\cal
R}=25.5$ and $K_s=20$~mag, and significant photometric measurements
somewhat fainter.  The galaxy number density is $1.3\times 10^5~{\rm
deg^{-2}}$ to ${\cal R}=25.0$~mag.  Galaxy number counts have slopes
$d\log N/dm=0.42$, $0.33$, $0.27$ and $0.31$ in the $U_n$, $G$, $\cal
R$ and $K_s$ bands, consistent with previous studies and the trend
that fainter galaxies are, on average, bluer.  Galaxy catalogs
selected in the $\cal R$ and $K_s$ bands are presented, containing
3607 and 488 sources, in field areas of $74.8$ and $59.4~{\rm
arcmin^2}$, to ${\cal R}=25.5$ and and $K_s=20$~mag.
\end{abstract}


\section{Introduction}

The Hubble Deep Field (HDF, Williams \etal\ 1996) was chosen to be at
high Galactic latitude, at low extinction, and free of bright or
unusual sources.  The Hubble Space Telescope (HST) images of the HDF
are the deepest optical images of the sky ever taken, reaching source
densities of roughly $10^6~{\rm deg^{-2}}$.  The HDF has quickly
become a ``standard field'' for the study of very faint extragalactic
sources; it has been studied at all wavelengths from x-ray to radio.
It is hoped that the huge multi-wavelength database which is
developing for this field will lead to a new understanding of faint
galaxy properties and evolution.  In this paper are presented the
catalogs of source fluxes and colors created for the Caltech Faint
Galaxy Redshift Survey, in a region of the sky centered on the the
HDF.

The Caltech Faint Galaxy Redshift Survey is a set of magnitude-limited
visual spectroscopic surveys, with visual- and near-infrared-selected
samples of very faint sources in blank fields.  The sources are
observed spectroscopically with the Low Resolution Imaging
Spectrograph (LRIS, Oke \etal\ 1995) instrument on the Keck Telescope.
The spectroscopic catalogs are presented in companion papers (Cohen
\etal\ 1999a, Cohen \etal\ 1999c).  Briefly, the survey comprises
several fields which together contain many hundreds of sources with
spectroscopically measured redshifts and multi-band photometry, down
to $R\approx 24.5$ or $K\approx 20$~mag, the current practical limit
of highly complete, magnitude-limited spectroscopic samples.  (Of
course the photometric catalogs, and some incomplete spectroscopic
samples, reach fainter fluxes than these.)  Early results include
studies of galaxy groups out to redshifts $z\approx 1$ (Cohen et al
1996a, 1996b, 1999b, 1999c) and measurements of broad-band and
emission line luminosity functions and their evolution to redshifts
$z\approx 1.5$ (Hogg \etal\ 1998a; Hogg 1998b; Cohen \etal\ 1999b).
The database of photometry and spectroscopy will be useful for studies
of faint galaxies and stars.

The HST images of the HDF are very small, covering only about
5~arcmin$^2$, so they are poorly matched to the 15~arcmin$^2$
spectroscopic field of the LRIS instrument.  For this reason the
spectroscopic surveys in the HDF are performed in a larger region of
the sky surrounding the HST image, with sources selected with the
ground-based data presented here.  The HDF observations with HST also
included short exposures (one or two orbits) for eight pointings
surrounding the HDF; these are referred to as the ``Flanking Fields''
(FF).  The potential for obtaining detailed morphological information
on the brighter sources at the resolution of HST therefore exists for
the photometric catalogs presented here, and the redshift catalogs
presented elsewhere (Cohen et al 1996a, 1996b, 1999c).

\section{Data}

For visual data, $U_n$, $G$ and $\cal R$ images taken with the COSMIC
camera (Kells et al 1998) at the prime focus of the 200-inch Hale
Telescope at the Palomar Observatory were used.  The COSMIC camera has
$0.283\times0.283~{\rm arcsec^2}$ pixels over a $9\times9~{\rm
arcmin^2}$ field of view.  The final, stacked images images are
$8.6\times 8.7~{\rm arcmin^2}$, centered on 12\,36\,51.4 +62\,13\,13
(J2000), ie, roughly centered on the HST image of the HDF (Williams
\etal\ 1996).  The visual images were taken in order to identify
candidate $z>3$ galaxies; details of the observations, calibration,
and reduction of these images are described in Steidel \etal\ (in
preparation).  The $U_n$, $G$ and $\cal R$ filters are described in
Steidel \& Hamilton (1993); briefly, they have effective wavelengths
of 3570, 4830 and 6930\,\AA, FWHM bandpasses of 700, 1200 and
1500\,\AA, and zero-magnitude flux densities of roughly 1550, 3890 and
2750~Jy.  These magnitudes are Vega-relative, not AB.

The southwest corner of the $\cal R$ image was contaminated by an
asteroid trail.  The trail was removed by transforming less sensitive
but higher-resolution Keck LRIS $R$-band images taken as set-up for
the spectroscopic program in this field (Cohen et al 1999c) onto the
same pixel scale, smoothing to the same seeing, and scaling to the
same zeropoint.  A strip of full width $4.7~{\rm arcsec}$ along the
straight trail was replaced with the smoothed, transformed Keck LRIS
image.  This thin strip of the $\cal R$-band image has slightly
higher-than-average noise.

For near-infrared data, an 8-arcmin diameter circular region centered
on the HDF was imaged on 1997 March 19--21 using a $K_s$ filter with a
near-infrared camera (Jarrett \etal\ 1994) mounted at the prime focus
of the 200-inch Hale Telescope.  The instrument reimages the focal
plane at 1:1 onto a NICMOS--3 $256 \times 256~{\rm pixel}$ HgCdTe
array (produced by Rockwell), producing a $0.494\times 0.494~{\rm
arcsec^2}$ projected pixel size and a $2.1\times2.1~{\rm arcmin^2}$
instantaneous field of view.  The $K_s$ filter has an effective
wavelength of $2.15~{\rm \mu m}$, a FWHM bandpass of $0.3~{\rm \mu
m}$, and a zero-magnitude flux density of roughly 708~Jy.  Fourteen
separate subfields, offset by 2~arcmin, were required in order to
mosaic the entire circular field; each of these subfield was imaged
once per night.  For each subfield each night, 45 separate frames were
taken; each frame consisted of six exposures of three seconds each,
coadded in the electronics before writing to disk.  The telescope was
dithered by 5--15~arcsec between frames.  As a result, each subfield
was exposed for 810~s each night, or 2430~s for the three nights.  The
seeing was $\sim 1.0$~arcsec FWHM for most of the three nights.  The
first two nights were judged photometric, and were calibrated using
the faint Solar-type standard stars of Persson et al (1998).

\begin{figure}
\plotone{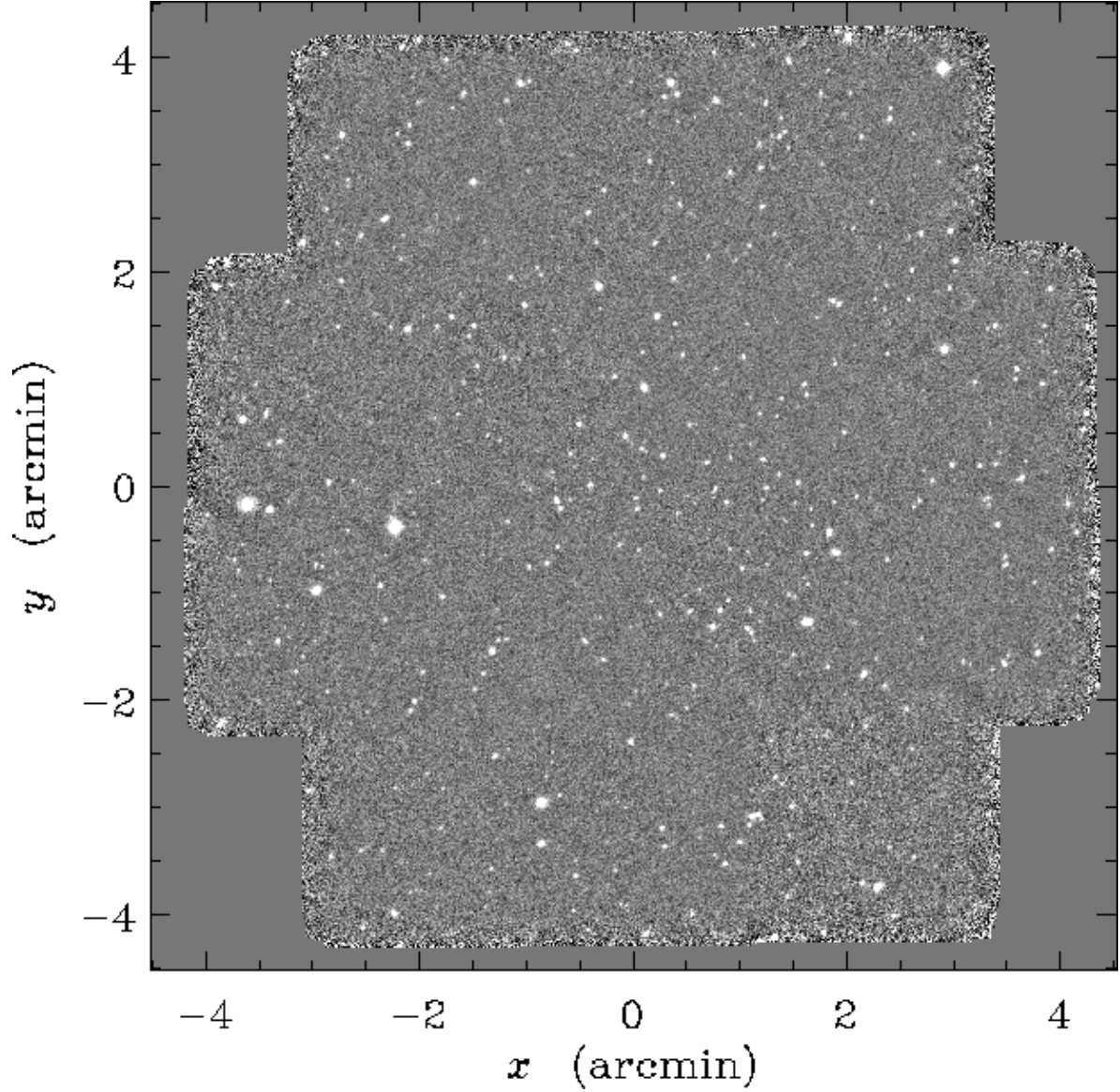}
\caption[The final $K_s$--band mosaic]{
The final $K_s$--band mosaic, stretched from $-5\,\sigma$ {\sl (black)}
to $5\,\sigma$ {\sl (white)}.}
\label{fig:Kimage}
\end{figure}

The $K_s$-band data were reduced by the method of Pahre \etal\ (1997).
Each subfield was reduced separately for each night.  The third
night's data were rescaled by factors of between 1.1 and 1.5 in order
to account for cirrus; the scaling factors were determined from a fit
to a large number of sources.  The subfields were then registered by
aligning the objects in common with adjacent subfields in the overlap
region.  Individual pixels in a given field were weighted by the
number of pointings contributing to that pixel.  A background level
was estimated at every pixel by median-filtering the mosaic with a
wide filter and sigma--clipping.  This background was subtracted in
order to remove subfield-to-subfield variations in the sky brightness
of the final mosaic.  The final $K_s$--band mosaic is displayed in
Figure~\ref{fig:Kimage}.

Table~\ref{tab:UGRK} gives the properties of the final, stacked
images.

\section{Source detection}

Sources were detected in all four images independently to construct
four catalogs, hereafter ``$U_n$-selected'', ``$G$-selected'', ``$\cal
R$-selected'' and ``$K_s$-selected''.  All catalogs were created with
the SExtractor source detection and photometry package (Bertin \&
Arnouts 1996).  The detection algorithm is as follows: Images are
smoothed with a Gaussian filter which has roughly the same FWHM as the
seeing (1.13~arcsec for the visual images and 1.5~arcsec for the
$K_s$-band image).  Sources in the smoothed image with central-pixel
surface brightness above a certain limit are added to the catalog.  If
a source has multiple peaks within its 1.2-$\sigma$ isophotal area on
the image (where $\sigma$ is the pixel-to-pixel root-mean-square
fluctuation in the sky brightness), each peak is split into a separate
catalog source if it contains at least one percent of the original
source's isophotal flux.

The $\cal R$-selected SExtractor catalog was augmented in two ways.
(1)~Several sources were added which, by eye, appear that they ought
to be split off of brighter nearby objects but were not.  These
sources, when above the $\cal R$-band flux limit, were added to the
$\cal R$-selected catalog directly.  (2)~Several very faint sources
were compiled into what is hereafter the ``supplemental'' catalog,
even though they are below the $\cal R$-selected catalog's flux limit,
because they have successful redshift measurements in the companion
paper Cohen et al (1999c).  The fluxes for the supplemental catalog are
all aperture magnitudes and the colors were measured as described
below.

The noise in the $K_s$ image is much worse along the edges of the
mosaic than at the center, which can lead to spurious detections.
Sources in the high-noise edges were removed from the $K_s$-selected
catalog, leaving a total area coverage of $59.4~{\rm arcmin^2}$.

\section{Calibration with HST imaging}

To maintain a flux or magnitude system consistent with previous work
in the HDF, the $U_n$, $G$ and $\cal R$ images are calibrated by
comparison with the extremely deep HST images of the HDF.  The
acquisition, reduction and calibration of the HST images are described
in Williams \etal\ (1996).  In what folows, the Vega-relative
calibrations of the HST images are used.

The absolute calibrations and effective wavelengths for the HST and
ground-based filters are used to compute the following transformation
equations under the assumption that the sources have roughly power-law
spectral energy distributions:
\begin{equation}
U_n= 0.53\, F300W + 0.47\, F450W - 0.29~{\rm mag}
\end{equation}
\begin{equation}
G= 0.82\, F450W + 0.18\, F606W - 0.07~{\rm mag}
\end{equation}
\begin{equation}
{\cal R}= 0.46\, F606W + 0.54\, F814W - 0.02~{\rm mag}
\end{equation}
where $F300W$, $F450W$, $F606W$ and $F814W$ are Vega-relative
magnitudes in the HST bandpasses of the same name.

\begin{figure}
\plotone{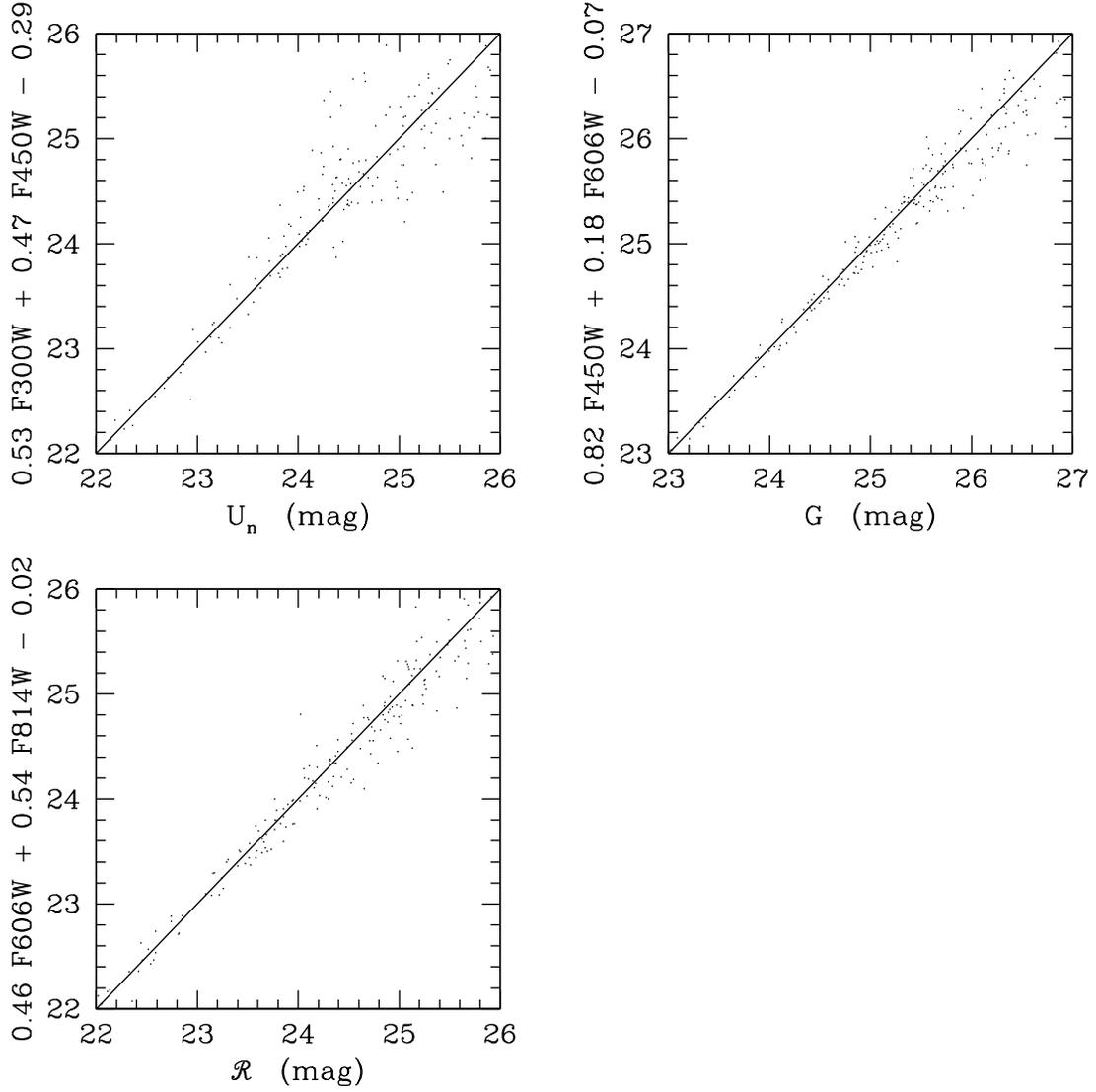}
\caption[Comparison of ground-based visual and transformed
HST magnitudes.]{Comparison of ground-based visual and transformed
HST magnitudes for the $\cal R$-selected sample.  These plots were
used to calibrate the $U_n$, $G$ and $\cal R$ images.}
\label{fig:checkphot}
\end{figure}

The ``Version 2'' HST HDF images (Williams \etal\ 1996) are
transformed onto the $U_n$, $G$ and $\cal R$ image coordinate system
and all seven images are Gaussian-smoothed to have the same effective
seeing.  Aperture magnitudes were measured for the $\cal R$-selected
sample through matched, 2-arcsec diameter apertures.  For calibration,
the Vega-relative magnitude zeropoints were used instead of the ``AB''
zeropoints used by Williams \etal\ (1996).  The measured $U_n$, $G$
and $\cal R$-band magnitudes are given zeropoints such that the
comparison with transformed HST magnitudes in
Figure~\ref{fig:checkphot} shows the best possible agreement.  This
HST-relative calibration ought to be good to roughly 5~percent.

\section{Photometry}

All catalog sources were photometered two ways: Isophotal magnitudes
were measured down to the 2-$\sigma$ isophote (where $\sigma$ is the
pixel-to-pixel root-mean-square fluctuation in the sky brightness).
Aperture magnitudes were measured through apertures of diameter
1.7~arcsec for the visual images and 2.0~arcsec for the $K_s$-band
image.  Corrections to account for flux outside the aperture were
added to the raw aperture magnitudes.  The aperture corrections were
measured from bright stars in the field and were found to be $-0.13$,
$-0.10$, $-0.10$ and $-0.12$~mag for the $U_n$, $G$, $\cal R$ and
$K_s$ images respectively.  These corrections correct aperture
magnitudes to total magnitudes for point sources; no adjustment was
made to account for galaxy size or extended structure in galaxies
because although faint galaxies are not point sources, in these
ground-based images there is almost no detectable difference between a
faint galaxy and star at the faintest levels.  Each source in the
catalogs was assigned a ``total magnitude'' which is the brighter of
the isophotal and corrected-aperture magnitudes.  In practice, this is
the isophotal magnitude for $79.9$~percent of sources to ${\cal
R}=24.5$~mag, and it is $42.5$~percent to ${\cal R}=25.5$~mag; and it
is $98.5$~percent to $K_s=19$~mag and $71.9$~percent to $K_s=20$~mag.
It should be noted that under this definition, the total magnitudes
are not expected to represent entire source fluxes, because there may
be significant flux at large radius and low surface-brightness around
these sources.  Unfortunately it is not possible to accurately measure
this low surface-brightness flux on a source-by-source basis.

\section{Color measurement}

To measure unbiased colors, the visual images were smoothed with
Gaussians to the same effective seeing as the $K_s$-band image.  A
catalog of over 500 objects common to the visual and $K_s$-band images
were used to derive the fourth-order polynomial transformation mapping
the visual images onto the $K_s$-band image and vice versa (with
NOAO/IRAF tasks ``geomap'' and ``geotran'').  Colors were measured
through matched apertures of diameter 2~arcsec.  For the $U_n$, $G$
and $\cal R$-selected catalogs, colors were measured in the smoothed
visual image and the $K_s$-band image transformed onto the visual
coordinates.  For the $K_s$-selected catalog, colors were measured in
the smoothed visual images transformed onto the $K_s$-band image
coordinates and the $K_s$-band image.

Color distributions for the four main catalogs are shown in
Figures~\ref{fig:Usel} through \ref{fig:Ksel}.  There are 1920 sources
with $U_n<25.0$~mag in the $U_n$-selected catalog, 2863 with
$G<26.0$~mag in the $G$-selected, 3607 with ${\cal R}<25.5$~mag in the
$\cal R$-selected, and 488 with $K_s<20.0$~mag in the $K_s$-selected.
The full $\cal R$-selected, $K_s$-selected, and supplemental catalogs
are given in Tables~\ref{tab:Rsel}, \ref{tab:Ksel}, and
\ref{tab:extras}.  {\it [For now the tables are available at \\
(http://www.sns.ias.edu/\~{}hogg/Hogg.Rsel.txt),
(http://www.sns.ias.edu/\~{}hogg/Hogg.Ksel.txt), and
(http://www.sns.ias.edu/\~{}hogg/Hogg.extras.txt).]}

\section{Astrometry}
\label{sec:astrometry}

Absolute positions were assigned to the $\cal R$-selected sources by
comparison with the Williams \etal\ (1996) and Phillips \etal\ (1997)
catalogs.  In the HST-imaged portion of the field, absolute positions
were found by identifying $\cal R$-selected sources with those in the
Williams \etal\ catalog.  In the flanking field, the $\sim 100$
sources in the Phillips \etal\ catalog were identified with $\cal
R$-selected sources.  A quadratic transformation was fit to the
relation between COSMIC pixel locations and absolute positions for the
identified sources.  This transformation was used to assign absolute
positions to all sources in the $\cal R$-selected catalog.  These
positions are given in Tables~\ref{tab:Rsel}, \ref{tab:Ksel}, and
\ref{tab:extras}.  Comparison with the radio maps of the HDF and
flanking fields (Richards et al 1998) shows that the absolute
positions have an rms accuracy of roughly 0.4~arcsec (Cohen et al
1999c).

\section{Completeness}
\label{sec:complete}

It appears from Figures~\ref{fig:Usel} through \ref{fig:Ksel} that the
catalogs are complete to roughly $U_n=25$, $G=26$, ${\cal R}=25.5$ and
and $K_s=20$~mag.  No completeness simulations have been performed
because the primary purpose of this study is to construct catalogs for
spectroscopy, not to measure ultra-deep number counts.  For the latter
study, better data exist and have been analyzed.  With typical colors,
objects with ${\cal R}>24$~mag and $K_s>20$~mag cannot routinely, or
with good completeness, be measured spectroscopically with the Keck
Telescope, so this catalog is appropriate for selection of a complete
spectroscopic sample.

\section{Discussion}

\begin{figure}
\plotone{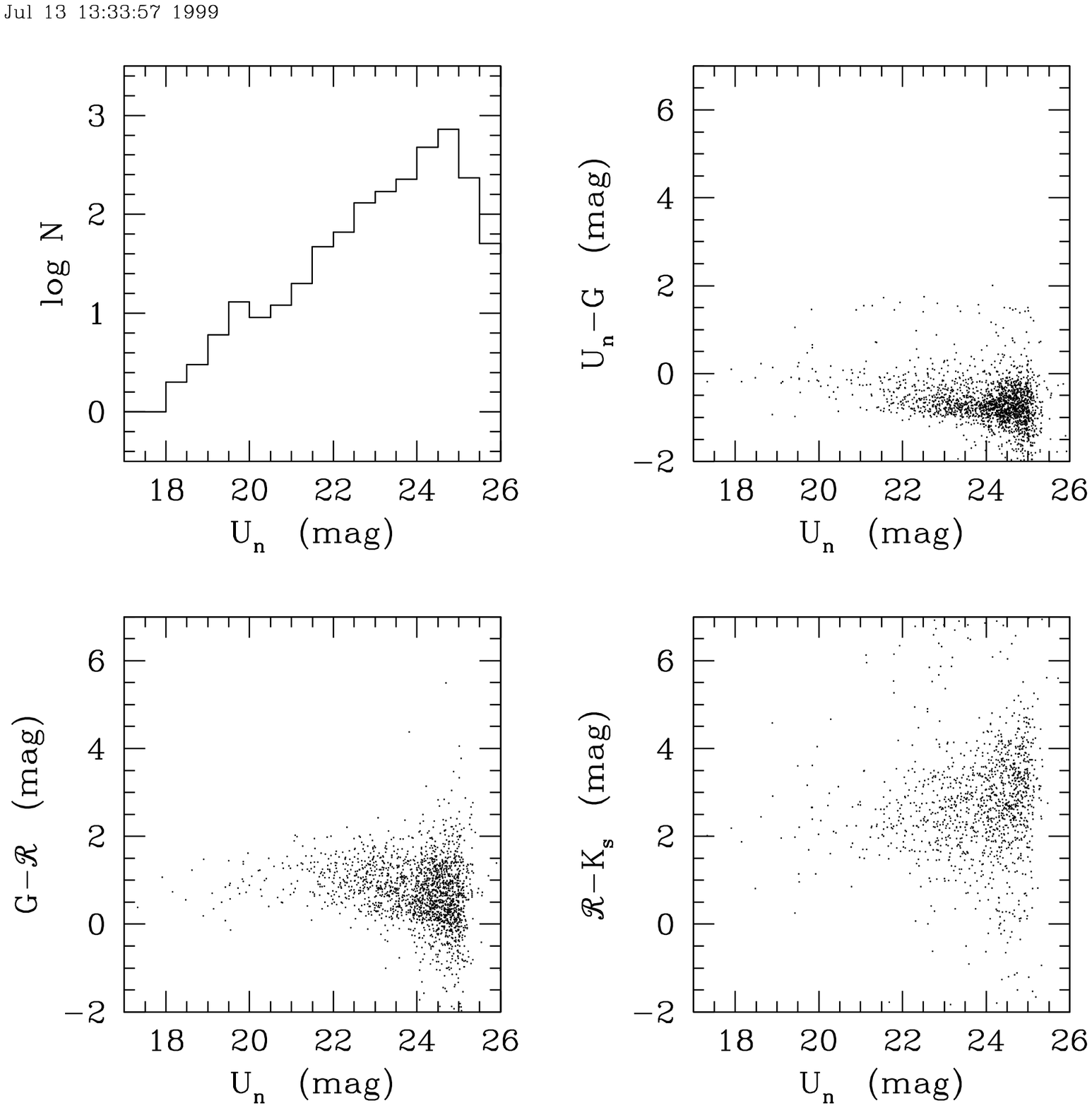}
\caption[$U_n$-selected sample numbers and colors.]{Number counts of
$U_n$-selected sources and their colors.  The number counts have not
been corrected for incompleteness, which appears to set in at
$U_n\approx 25$~mag.}
\label{fig:Usel}
\end{figure}

\begin{figure}
\plotone{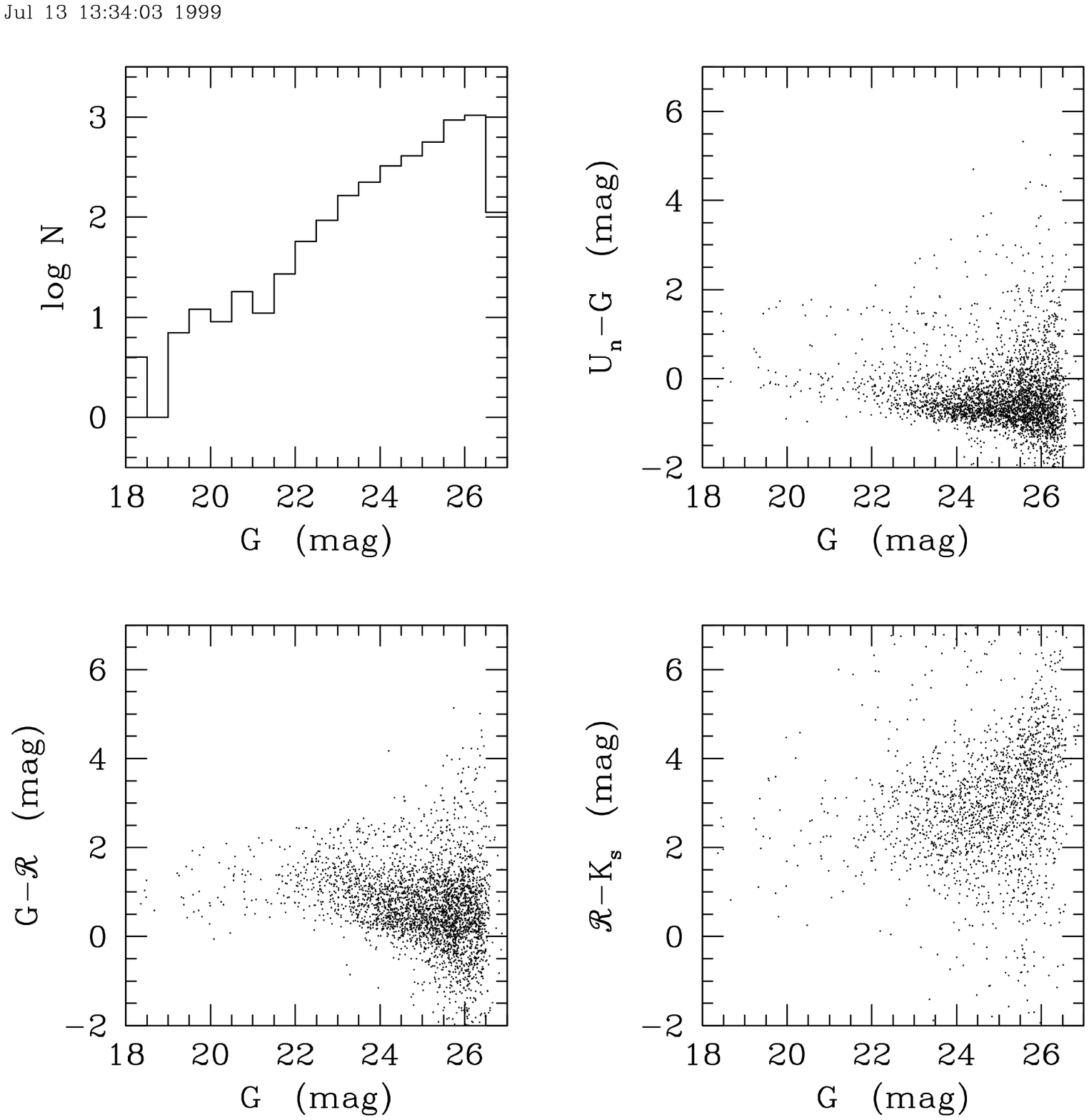}
\caption[$G$-selected sample numbers and colors.]{Number counts of
$G$-selected sources and their colors.  The number counts have not
been corrected for incompleteness, which appears to set in at
$G\approx 26$~mag.}
\label{fig:Gsel}
\end{figure}

\begin{figure}
\plotone{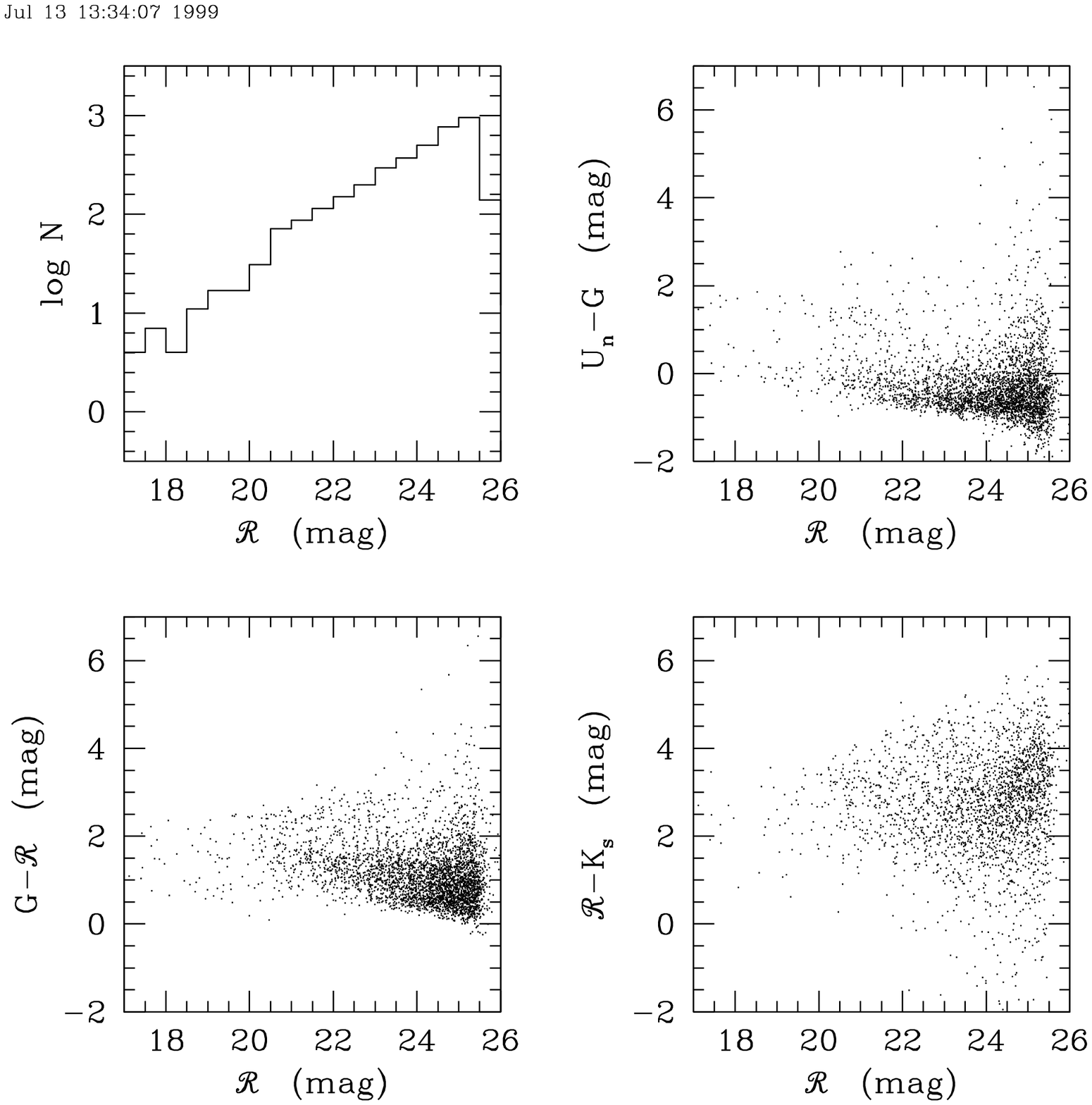}
\caption[$\cal R$-selected sample numbers and colors.]{Number counts of
$\cal R$-selected sources and their colors.  The number counts have not
been corrected for incompleteness, which appears to set in at
${\cal R}\approx 25.5$~mag.}
\label{fig:Rsel}
\end{figure}

\begin{figure}
\plotone{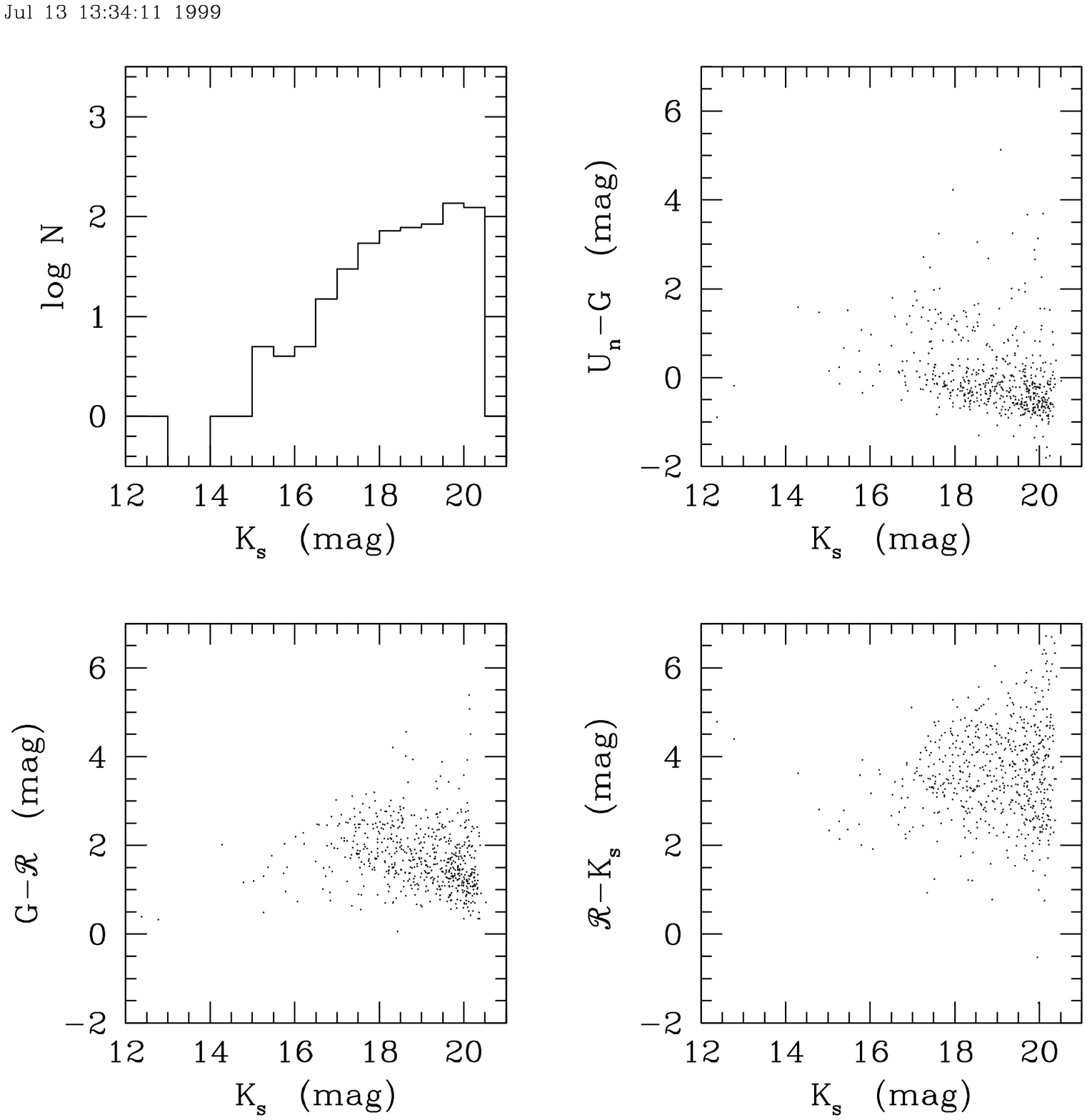}
\caption[$K_s$-selected sample numbers and colors.]{Number counts of
$K_s$-selected sources and their colors.  The number counts have not
been corrected for incompleteness, which appears to set in at
$K_s\approx 20$~mag.}
\label{fig:Ksel}
\end{figure}

The results of this survey are entirely contained in
Figures~\ref{fig:Usel} through \ref{fig:Ksel}.  However, they can be
compared with the results of other authors.  When divided by the solid
angle of the $8.6\times8.7~{\rm arcmin^{2}}$ field, the integrated
number of sources is $1.3\times 10^5~{\rm deg^{-2}}$ to ${\cal
R}=25.0$~mag.  This is consistent with number counts from similar
studies (\eg, Hogg \etal\ 1997b).  The color distributions are also
consistent with the results of previous studies, in mean and scatter
(Hogg \etal\ 1997a, 1997b; Pahre \etal\ 1997).

Number--flux relations of the power-law form $d\log N/dm=Q$, where $Q$
is a constant, can be fit to the $U_n$, $G$ and $\cal R$-selected
catalogs over the 4-magnitude range terminating at the completeness
limits given in Section~\ref{sec:complete}.  In the $K_s$-selected
catalog the fit is performed only over $18<K<20$~mag because many
studies have shown that the slope changes significantly at $K\approx
18$~mag (\eg, Gardner et al 1993; Djorgovski \etal\ 1995).  The
resulting faint-end slopes are $Q=0.42$, $0.33$, $0.27$ and $0.31$ for
the $U_n$, $G$, $\cal R$ and $K_s$ counts respsectively.  These slopes
are consistent with those found in previous studies (Djorgovski \etal\
1995; Metcalfe \etal\ 1995; Hogg \etal\ 1997b; Pahre \etal\ 1997).

Although all these observations are consistent with the results of
previous observational studies, the bulk of the faint sources are
significantly bluer than normal, bright galaxies would be if there
were no evolution in galaxy spectra.  For example, a non-evolving
spiral galaxy would have ${\cal R}-K_s\approx 3$~mag at redshift
$z=0.6$, and the bluest local galaxies would have ${\cal R}-K_s\approx
2.5$~mag, but in the samples presented here, where the median redshift
is roughly 0.6 (Cohen \etal\ 1999a, 1999c), there are many galaxies
with ${\cal R}-K_s<2$~mag.  The appearance of this extremely blue
population in faint samples is a consequence of the high star
formation rates at intermediate and high redshift relative to those of
in the present-day Universe, as inferred from metallicity in
Lyman-alpha clouds (Pei \& Fall 1995), ultraviolet luminosity density
(Lilly et al 1996; Connolly et al 1997; Madau et al 1998) and emission
line strengths (Hammer et al 1997; Heyl et al 1997; Small et al 1997;
Hogg et al 1998).  This evolutionary effect, the decrease in star
formation rate since redshift unity, is perhaps the most widely and
independently confirmed result in the study of field galaxy evolution.

\acknowledgements The Hubble Deep Field (HDF) database was planned,
taken, reduced and made public by a large team at Space Telescope
Science Institute headed by Bob Williams.  We thank the referee, Alan
Dressler, for timely and helpful criticism.  This study is based on
observations made at the Palomar Observatory, which is owned and
operated by the California Institute of Technology; and with the
NASA/ESA Hubble Space Telescope, which is operated by AURA under NASA
contract NAS 5-26555.  Primary financial support was provided under
NSF grant AST~95-29170.  Some additional support was provided by
Hubble Fellowship grants HF-01093.01-97A and HF-01099.01-97A from
STScI, which is operated by AURA under NASA contract NAS~5-26555.


\clearpage
\begin{deluxetable}{lccccc}
\tablewidth{0pt}
\tablecaption{
        Imaging data parameters, for final, stacked mosaics
	\label{tab:UGRK}
}
\tablehead{
   \colhead{band}
 & \colhead{solid angle}
 & \colhead{exposure}
 & \colhead{pixel size}
 & \colhead{seeing FWHM}
 & \colhead{$1\,\sigma$ in aperture\tablenotemark{a}}
\\
   \colhead{}
 & \colhead{(arcmin$^2$)}
 & \colhead{(s)}
 & \colhead{(arcsec)}
 & \colhead{(arcsec)}
 & \colhead{(mag)}
}
\startdata
$U_n$    & 75 & 23400 & 0.283 & 1.3 & 26.9 \nl
$G$      & 75 &  7200 & 0.283 & 1.2 & 28.1 \nl
$\cal R$ & 75 &  6000 & 0.283 & 1.1 & 27.2 \nl
$K_s$    & 59 &  2430\tablenotemark{b} & 0.494 & 1.5 & 22.1 \nl
\enddata
\tablenotetext{a}{The $1\,\sigma$ magnitude is the flux corresponding
to a $1\,\sigma$ variation in the sky in a 2~arcsec diameter
focal-plane aperture, the aperture used for color measurements.}
\tablenotetext{b}{This exposure time applies to each of the 14
separate subfields.}
\end{deluxetable}

\clearpage
\begin{deluxetable}{c}
\tablecaption{
        The $\cal R$-selected catalog\tablenotemark{a}
	\label{tab:Rsel}
}
\tablehead{}
\startdata
\enddata
\tablenotetext{a}{For now, the contents of this table can be found at
(http://www.sns.ias.edu/\~{}hogg/Hogg.Rsel.txt)}
\end{deluxetable}

\begin{deluxetable}{c}
\tablecaption{
        The $K_s$-selected catalog\tablenotemark{a}
	\label{tab:Ksel}
}
\tablehead{}
\startdata
\enddata
\tablenotetext{a}{For now, the contents of this table can be found at
(http://www.sns.ias.edu/\~{}hogg/Hogg.Ksel.txt)}
\end{deluxetable}

\begin{deluxetable}{c}
\tablecaption{
        The supplemental catalog\tablenotemark{a}
	\label{tab:extras}
}
\tablehead{}
\startdata
\enddata
\tablenotetext{a}{For now, the contents of this table can be found at
(http://www.sns.ias.edu/\~{}hogg/Hogg.extras.txt)}
\end{deluxetable}

\clearpage
\begin{figure}
\plotone{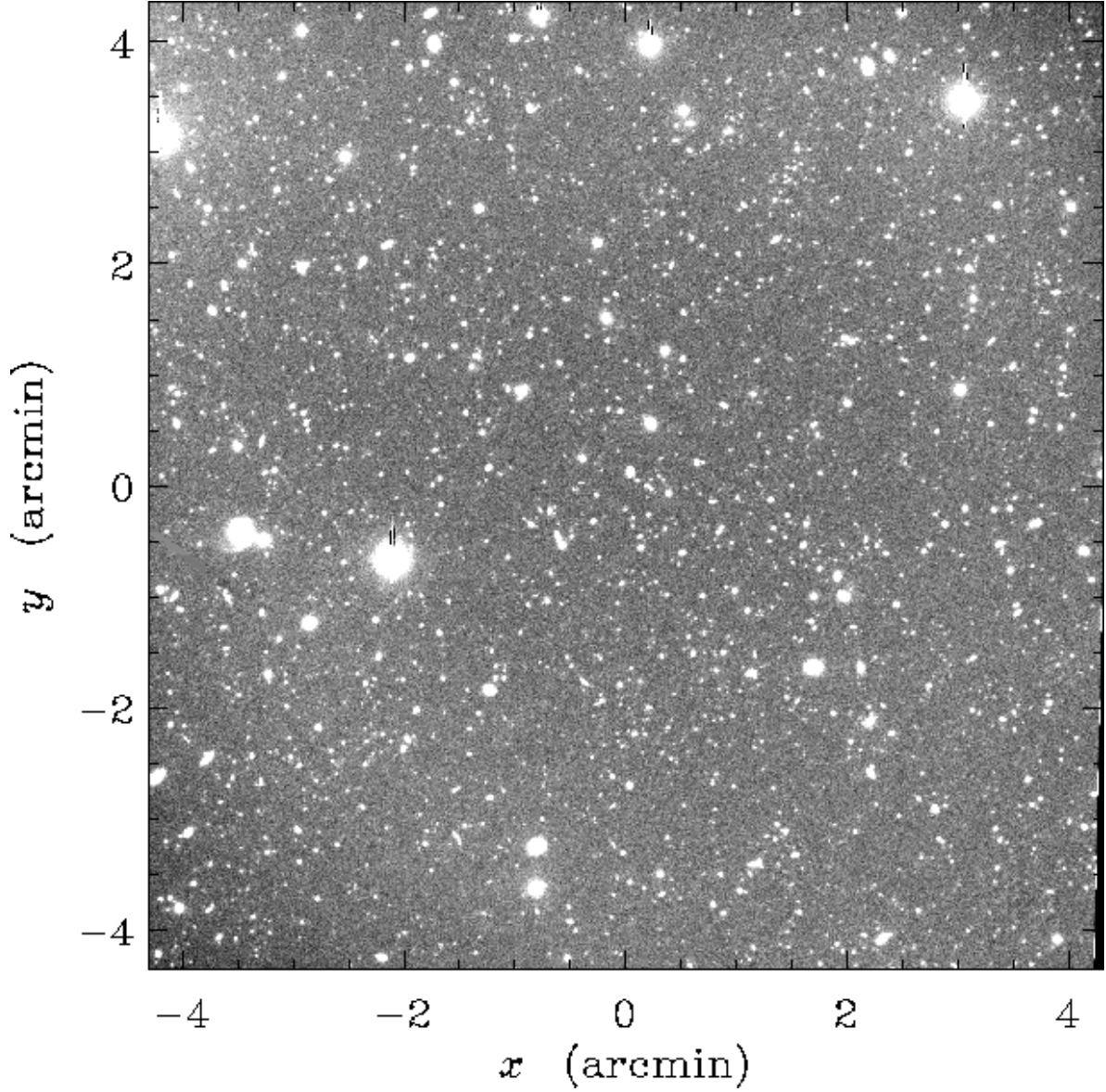}
\caption[The final $\cal R$--band mosaic]{
The final $\cal R$--band mosaic, stretched from $-5\,\sigma$ {\sl (black)}
to $5\,\sigma$ {\sl (white)}.  The field center is 12\,36\,51.4
+62\,13\,13 (J2000).}
\label{fig:Rimage}
\end{figure}

\end{document}